# Knightian Robustness of Single-Parameter Domains

Alessandro Chiesa     Silvio Micali     Zeyuan Allen Zhu

MIT

March 25, 2014

**Abstract**

We consider players that have very limited knowledge about their own valuations. Specifically, the only information that a *Knightian player i* has about the profile of true valuations, $\theta^*$, consists of a set of distributions, from one of which $\theta_i^*$ has been drawn.

We prove a "robustness" theorem for Knightian players in single-parameter domains: every mechanism that is weakly dominant-strategy truthful for classical players continues to be well-behaved for Knightian players that choose undominated strategies.

# 1 Introduction

In [CMZ14] we motivate the problem of mechanism design for Knightian players, and prove that (1) dominant-strategy mechanisms for single-good and multi-unit auctions cannot provide good social-welfare efficiency, but (2) the second-price and Vickrey mechanisms deliver good social-welfare performance, for these two settings, in undominated strategies.

In this report, we prove a "robustness" theorem for single-parameter domains. Namely, consider a mechanism $M$ for a single-parameter domain and suppose that $M$, when players have perfect information about their own valuations, is weakly dominant-strategy truthful. Now consider the same mechanism $M$, but with Knightian players that, not having any dominant strategy to play, choose to play undominated strategies. We prove that the set of undominated strategies is well-behaved, in the sense that these strategies do not deviate from the players' approximate information about his own valuation.

# 2 Model

In a classical single-parameter domain, there is a set $\mathcal{A}$, the set of all possible allocations; for each player $i$ there exists a publicly known subset $\mathcal{S}_i \subseteq \mathcal{A}$; and the set of possible valuations for player $i$, $\Theta_i$, consists of all functions mapping $\mathcal{A}$ to the reals, subject to the following constraints: for each $\theta_i \in \Theta_i$,

(1) $\theta_i(x) = 0 \quad \forall x \notin \mathcal{S}_i$ and

(2) $\theta_i(x) = \theta_i(y) \quad \forall x, y \in \mathcal{S}_i$.

We denote the true valuation of player $i$ by $\theta_i^*$.

(The term "single-parameter" derives from the fact that each $\theta_i \in \Theta_i$ coincides with a single number: $i$'s value for, say, the lexicographically first element of $\mathcal{S}_i$. The term "classical" emphasizes that each player knows exactly his own true valuation.)

The set of possible outcomes is $\Omega \stackrel{\text{def}}{=} \mathcal{A} \times \mathbb{R}^n_{\geq 0}$. If $(A, P) \in \Omega$, we refer $P_i$ as the price charged to player $i$. We assume quasi-linear utilities. That is, the utility function $U_i$ of a player $i$ maps a valuation $\theta_i$ and an outcome $\omega = (A, P)$ to $U_i(\theta_i, \omega) \stackrel{\text{def}}{=} \theta_i(A) - P_i$.

If $\omega$ is a distribution over outcomes, we also denote by $U_i(\theta_i, \omega)$ the expected utility of player $i$.

Single-parameter domains are general enough to include several settings of interest: in particular, provision of a public good[1] [Cla71], bilateral trades [MS83], and buying a path in a network [NR01].

## 2.1 Knightian Valuation Uncertainty

In our model, a player $i$'s sole information about $\theta^*$ consists of $\mathcal{K}_i$, a set of distributions over $\Theta_i$, from one of which $\theta_i^*$ has been drawn. (The true valuations are uncorrelated.) That is, $\mathcal{K}_i$ is $i$'s sole (and private) information about his own true valuation $\theta_i^*$. Furthermore, for every opponent $j$, $i$ has no information (or beliefs) about $\theta_j^*$ or $\mathcal{K}_j$.

Given that all he cares about is his expected (quasi-linear) utility, a player $i$ may 'collapse' each distribution $D_i \in \mathcal{K}_i$ to its expectation $\mathbb{E}_{\theta_i \sim D_i}[\theta_i]$.[2] Therefore, for single-parameter domains, a *mathematically equivalent* formulation of the Knightian valuation model is the following:

**Definition 2.1** (Knightian valuation model). *For each player $i$, $i$'s sole information about $\theta^*$ is a set $K_i$, the* candidate (valuation) set *of $i$, such that $\theta_i^* \in K_i \subset \Theta_i$.*

*We refer to an element of $K_i$ as a* candidate valuation.

In Knightian valuation model, a mechanism's performance will of course depend on the inaccuracy of the players' candidate sets, which we measure as follows.

**Definition 2.2.** *Let $K_i^\perp \stackrel{\text{def}}{=} \inf K_i$ and $K_i^\top \stackrel{\text{def}}{=} \sup K_i$.*

*The candidate set $K_i$ of a player $i$ is* (at most) $\delta$-approximate *if $K_i^\top - K_i^\perp \leq \delta$.*

*A single-parameter domain is* (at most) $\delta$-approximate *if each $K_i$ is $\delta$-approximate.*

---

[1] Indeed, in the provision of a public good, $\mathcal{A}$ has just two elements, $a$ (i.e., the good is provided), which different players may value differently, and $b$ (i.e., the good is not provided), which all players value 0.

[2] Whatever the auction mechanism used, this equivalence holds for any auction where each $\Theta_i$ is a *convex* set. In particular, this includes unrestricted combinatorial auctions of $m$ distinct goods.

## 2.2 Social Welfare, Mechanisms, and Knightian Dominance

**Social welfare.** The social welfare of an allocation $A \in \mathcal{A}$, $\mathrm{SW}(A)$, is defined to be $\sum_i \theta_i^*(A)$; and the maximum social welfare, MSW, is defined to be $\max_{A \in \mathcal{A}} \mathrm{SW}(A)$. (That is, SW and MSW continue to be defined relative to the players' true valuations $\theta_i^*$, whether or not the players know them exactly.)

More generally, the social welfare of an allocation $A$ relative to a valuation profile $\theta$, $\mathrm{SW}(\theta, A)$, is $\sum_i \theta_i(A)$; and the maximum social welfare relative to $\theta$, $\mathrm{MSW}(\theta)$, is $\max_{A \in \mathcal{A}} \mathrm{SW}(\theta, A)$. Thus, $\mathrm{SW}(A) = \mathrm{SW}(\theta^*, A)$ and $\mathrm{MSW} = \mathrm{MSW}(\theta^*)$.

**General mechanisms and strategies.** A mechanism $M$ specifies, for each player $i$, a set $S_i$. We interchangeably refer to each member of $S_i$ as a pure *strategy/action/report* of $i$, and similarly, a member of $\Delta(S_i)$ a mixed strategy/action/report of $i$.

After each player $i$, simultaneously with his opponents, reports a strategy $s_i$ in $S_i$, $M$ maps the reported strategy profile $s$ to an outcome $M(s) \in \Omega$.

If $M$ is probabilistic, then $M(s) \in \Delta(\Omega)$. Thus, as per our notation, $U_i(\theta_i, M(s)) \stackrel{\text{def}}{=} \mathbb{E}_{\omega \sim M(s)}[U_i(\theta_i, \omega)]$ for each player $i$.

Note that $S_i = \Theta_i$ for the direct mechanisms in the classical setting, but may be arbitrary in general.

**Knightian undominated strategies.** Given a mechanism $M$, a pure strategy $s_i$ of a player $i$ with a candidate set $K_i$ is *(weakly) undominated*, in symbols $s_i \in \mathsf{UD}_i(K_i)$, if $i$ does not have another (possibly mixed) strategy $\sigma_i$ such that

(1) $\forall \theta_i \in K_i \ \forall s_{-i} \in S_{-i} \quad \mathbb{E} U_i(\theta_i, M(\sigma_i, s_{-i})) \geq U_i(\theta_i, M(s_i, s_{-i}))$, and

(2) $\exists \theta_i \in K_i \ \exists s_{-i} \in S_{-i} \quad \mathbb{E} U_i(\theta_i, M(\sigma_i, s_{-i})) > U_i(\theta_i, M(s_i, s_{-i}))$.

If $K$ is a product or a profile of candidate sets, that is, if $K = (K_1, \ldots, K_n)$ or $K = K_1 \times \cdots \times K_n$, then $\mathsf{UD}(K) \stackrel{\text{def}}{=} \mathsf{UD}_1(K_1) \times \cdots \times \mathsf{UD}_n(K_n)$.

Note that the above notion of an undominated strategy is a natural extension of its classical counterpart, but other extensions are possible.

**Weakly dominant-strategy truthfulness in classical settings.** Finally, let us recall what it means for a mechanism $M$ to be weakly dominant-strategy truthful

(weakly DST) when every player $i$ knows $\theta_i^*$ exactly. Namely, for each player $i$:

(0)    $S_i = \Theta_i$

(1)    $\forall v_i \in \Theta_i \ \forall v_i' \in \Theta_i \ \forall v_{-i} \in \Theta_{-i}$      $U_i\big(v_i, M(v_i, v_{-i})\big) \geq U_i\big(v_i, M(v_i', v_{-i})\big)$

(2)    $\forall v_i \in \Theta_i \ \forall v_i' \in \Theta_i \setminus \{v_i\} \ \exists v_{-i} \in \Theta_{-i}$    $U_i\big(v_i, M(v_i, v_{-i})\big) > U_i\big(v_i, M(v_i', v_{-i})\big)$ .

(For comparison, the notion of a DST mechanism omits the last condition above.)

## 3   Result

We prove the Knightian robustness of many mechanisms at once as follows.

> **Theorem 1.** *Let $M$ be a weakly dominant-strategy truthful mechanism for classical single-parameter domains. Then, in this domain with Knightian valuation uncertainty, for every player $i$, $\mathsf{UD}(K_i) \subseteq \big[K_i^\perp, K_i^\top\big]$.*

**Discussion.** The above theorem implies that the behavior of (weakly dominant-strategy truthful) mechanisms in a $\delta$-approximate single-parameter domains gracefully degrades with $\delta$. In particular, it implies that, when applied to the provision of a public good in the presence of $n$ Knightian players, the VCG mechanism guarantees, in undominated strategies, a social welfare $\geq \mathsf{MSW} - 2n\delta$. As another example, when applied to buying paths in a network, the VCG mechanism guarantees a social welfare $\geq \mathsf{MSW} - 2m\delta$, where $m$ is the number of edges in the network. Finally, we note that the proof of Theorem 1 easily extends to imply an analogous result for the VCG mechanism for *single-minded combinatorial auctions*, which are not quite single-parameter domains.[3]

More generally, Theorem 1 implies that, for all weakly dominant-strategy mechanisms $M$ (which include those of [Cla71, MS83, NR01])

*'the outcome $M(v)$ is sufficiently good*

*whenever $\max_i |v_i - \theta_i^*|$ is sufficiently small for all $i$ and $\theta_i^* \in K_i$'.*

---

[3] In such an auction, there are $m$ distinct goods, and each player $i$ values, positively and for the same amount $\theta_i^*$, only the supersets of a given subset $\mathcal{S}_i$ of the goods. This auction is not single-parameter because $\mathcal{S}_i$ is *private*, that is, known solely to $i$. Accordingly, $i$'s true valuation can be fully described only by the number $\theta_i^*$ *and* the subset $\mathcal{S}_i$. The VCG mechanism for single-minded auctions ensures, in undominated strategies, a social welfare that is at least $\mathsf{MSW} - 2\min\{n, m\}\delta$.

**Proof.** The theorem is obvious when $K_i = \{\theta_i^*\}$ is a singleton: since reporting the truth is a weakly dominant strategy, it dominates all other strategies so that $\mathsf{UD}(K_i) = \{\theta_i^*\}$ must also be a singleton. For the rest of the proof we assume that $K_i$ has at least two distinct valuations.

We begin by recalling the following fact about dominant-strategy truthful mechanisms in single-parameter domains where each player perfectly knows his own true valuation [AT01]:

> Let $M$ be a mechanism for a single-parameter domain, and let $f_i(v) \in [0,1]$ be the probability that the allocation chosen by $M$, under strategy profile $v$, is in player $i$'s set $\mathcal{S}_i$. Then, $M$ is dominant-strategy truthful if and only if (a) $f$ is monotonically non-decreasing, i.e., $f_i(v_i, v_{-i}) \leq f_i(v_i', v_{-i})$ whenever $v_i \leq v_i'$, and (b) player $i$'s expected price on input $v$, denoted by $p_i(v)$, equals to $v_i \cdot f_i(v_i, v_{-i}) - \int_0^{v_i} f_i(z, v_{-i})\, dz$.

Having recalled the above fact, we now prove that, for any Knightian player $i$ with candidate set $K_i = [K_i^\perp, K_i^\top]$,
$$v_i \in \mathsf{UD}_i(K_i) \implies v_i \in [K_i^\perp, K_i^\top].$$
Let $v_i^\perp \stackrel{\text{def}}{=} K_i^\perp$ and $v_i^\top \stackrel{\text{def}}{=} K_i^\top$, and consider any strategy $v_i \in \mathsf{UD}_i(K_i)$. If $v_i \in K_i = [v_i^\perp, v_i^\top]$ then we are done. Otherwise, suppose that $v_i < v_i^\perp$. (The other case, $v_i > v_i^\top$, can be shown analogously.)

We first claim that, for player $i$, reporting $v_i^\perp$ is no worse than reporting $v_i$. Indeed, fixing any (pure) strategy sup-profile $v_{-i}$ for the other players and any possible true valuation $\theta_i \in K_i$, and letting $v^\perp = (v_i^\perp, v_{-i})$ and $v = (v_i, v_{-i})$, we compute that

$$\mathbb{E}\big[U_i\big(\theta_i, M(v^\perp)\big)\big] - \mathbb{E}\big[U_i\big(\theta_i, M(v)\big)\big]$$
$$= \big(f_i(v^\perp) - f_i(v)\big) \cdot \theta_i - \big(p_i(v^\perp) - p_i(v)\big)$$
$$= \big(f_i(v^\perp) - f_i(v)\big) \cdot \theta_i - \left(v_i^\perp \cdot f_i(v^\perp) - \int_0^{v_i^\perp} f_i(z, v_{-i})\, dz - v_i \cdot f_i(v) + \int_0^{v_i} f_i(z, v_{-i})\, dz\right)$$
$$= \big(f_i(v^\perp) - f_i(v)\big) \cdot (\theta_i - v_i^\perp) + \int_{v_i}^{v_i^\perp} \big(f_i(z, v_{-i}) - f_i(v)\big)\, dz\ .$$

Now note that $\theta_i \in K_i$ implies that $\theta_i - v_i^\perp = \theta_i - K_i^\perp \geq 0$. Moreover, by the monotonicity of $f$, whenever $z \geq v_i$, it holds that $f_i(z, v_{-i}) \geq f_i(v)$. Therefore we deduce that the above difference is greater than or equal to zero. We conclude that reporting $v_i^\perp$ is no worse than reporting $v_i$.

Next there are two subcases. If $\mathbb{E}\big[U_i\big(\theta_i, M(v^\perp)\big)\big] - \mathbb{E}\big[U_i\big(\theta_i, M(v)\big)\big]$ equals to zero for all $\theta_i \in K_i$ and for all $v_{-i}$, then, using the fact that $K_i$ has at least two distinct valuations, we conclude that for $i$, the allocation probability and (expected) price in outcomes $M(v_i, v_{-i})$ and $M(v_i^\perp, v_{-i})$ are the same, independent of $v_{-i}$. This contradicts the fact that $M$ is weakly dominant-strategy truthful in the classical setting, since $U_i(v_i, M(v_i, v_{-i}))$ must be strictly greater than $U_i(v_i, M(v_i^\perp, v_{-i}))$ at least for some $v_{-i}$.

Otherwise, if there exist some $\theta_i^*$ and some $v_{-i}^*$ that make the difference $\mathbb{E}\big[U_i\big(\theta_i, M(v^\perp)\big)\big] - \mathbb{E}\big[U_i\big(\theta_i, M(v)\big)\big]$ non-zero, it must follow that the difference is strictly positive. For such $\theta_i^*$ and $v_{-i}^*$, reporting $v_i^\perp$ is therefore strictly better than reporting $v_i$, so by definition $v_i^\perp$ weakly dominates $v_i$ for player $i$, leading to a contradiction to $v_i \in \mathsf{UD}_i(K_i)$.

This concludes the proof of Theorem 1. ∎

# References

[AT01] Aaron Archer and Éva Tardos. Truthful mechanisms for one-parameter agents. In *Proceedings of the 2001 IEEE 42nd Annual Symposium on Foundations of Computer Science*, FOCS '01, pages 482–491. IEEE Computer Society, 2001.

[Cla71] Edward H. Clarke. Multipart pricing of public goods. *Public Choice*, 11:17–33, 1971.

[CMZ14] Alessandro Chiesa, Silvio Micali, and Zeyuan Allen Zhu. Knightian robustness of the Vickrey mechanism. *ArXiv e-prints*, abs/xxxx.xxxx, March 2014. to appear.

[MS83] Roger B Myerson and Mark A Satterthwaite. Efficient mechanisms for bilateral trading. *Journal of economic theory*, 29(2):265–281, 1983.

[NR01] Noam Nisan and Amir Ronen. Algorithmic mechanism design. *Games and Economic Behavior*, 35:166–196, 2001.